\documentclass[twocolumn]{aastex631}

\usepackage{amssymb,amsmath,framed,amsthm}
\usepackage{graphicx}
\usepackage{epstopdf, epsfig}
\hypersetup{citecolor=blue,filecolor=blue,linkcolor=blue,urlcolor=blue}
\usepackage[T1]{fontenc}
\usepackage[whole]{bxcjkjatype} 

\makeatletter
\@ifpackageloaded{stix} 
{
}{                      
	\usepackage{mathrsfs}
  \usepackage{txfonts}          

}
\ifx\bm\undefined
  \newcommand{\bm}[1]{{\mathbf{#1}}} 
\else
  \renewcommand{\bm}[1]{{\mathbf{#1}}} 
\fi
\makeatother

\newcommand{\p}{\partial}
\newcommand{\f}[2]{\frac{#1}{#2}}
\newcommand{\mr}[1]{\mathrm{#1}}
\newcommand{\lf}{\left}
\newcommand{\ri}{\right}

\newcommand{\dd}[2]{\frac{\rmd#1}{\rmd#2}}
\newcommand{\pp}[2]{\frac{\p #1}{\p #2}}

\newcommand{\Div}{\nabla\cdot}
\newcommand{\Curl}{\nabla\times}
\newcommand{\nbl}{\nabla}

\newcommand{\+}{{\perp}}
\newcommand{\unit}[1]{\hat{\bm{#1}}}

\makeatletter
\newcommand{\vast}{\bBigg@{4}}
\newcommand{\Vast}{\bBigg@{5}}

\newcommand{\zhat}{\unit{z}}

\newcommand{\calliope}{\textsc{\textsf{Calliope}}}
\makeatother

\newcommand{\calD}{\mathcal{D}}

\newcommand{\calL}{\mathcal{L}}

\newcommand{\calN}{\mathcal{N}}

\DeclareMathAlphabet{\mathpzc}{OT1}{pzc}{m}{it}

\newcommand{\rmd}{\mathrm{d}}

\newcommand{\rmf}{\mathrm{f}}

\newcommand{\rmi}{\mathrm{i}}

\newcommand{\rms}{\mathrm{s}}

\newcommand{\rmA}{\mathrm{A}}

\newcommand{\rmS}{\mathrm{S}}

\shorttitle{Pseudospectral shearing MHD code with a pencil decomposition}
\shortauthors{Y. Kawazura}

\begin{document}

\title{\calliope: Pseudospectral shearing magnetohydrodynamics code with a pencil decomposition}

\correspondingauthor{Y. Kawazura}
\email{kawazura@tohoku.ac.jp}

\author[0000-0002-8787-5170]{Y. Kawazura}
\affiliation{Frontier Research Institute for Interdisciplinary Sciences, Tohoku University, 6-3 Aoba, Aramaki, Aoba-ku, Sendai 980-8578, Japan}
\affiliation{Department of Geophysics, Graduate School of Science, Tohoku University, 6-3 Aoba, Aramaki, Aoba-ku, Sendai 980-8578, Japan}
\affiliation{Astrophysical Big Bang Laboratory, RIKEN, 2-1 Hirosawa, Wako, Saitama 351-0198, Japan}



\begin{abstract}
The pseudospectral method is a highly accurate numerical scheme suitable for turbulence simulations. 
We have developed an open-source pseudospectral code, \calliope, which adopts the P3DFFT library~\citep{Pekurovsky2012} to perform a fast Fourier transform with the two-dimensional (pencil) decomposition of numerical grids. 
\calliope~can solve incompressible magnetohydrodynamics (MHD), isothermal compressible MHD, and rotational reduced MHD with parallel computation using very large numbers of cores ($> 10^5$ cores for $2048^3$ grids). 
The code can also solve for local magnetorotational turbulence in a shearing frame using the remapping method~\citep{Rogallo1981,Umurhan2004}. 
\calliope~is currently the only pseudospectral code that can compute magnetorotational turbulence using pencil-domain decomposition. 
This paper presents the numerical scheme of \calliope~and the results of linear and nonlinear numerical tests, including compressible local magnetorotational turbulence with the largest grid number reported to date.
\end{abstract}



\section{Introduction} \label{s:introduction}
Accretion disks are ubiquitous in the universe, including active galactic nuclei, close binary systems, and protostars. 
Thus, accretion disks have long been a subject of interest to astrophysicists. 
For mass accretion to occur, the accretion disk must be turbulent. 
The origin of the turbulence is believed to be due to magnetorotational instability~\citep[MRI; ][]{Balbus1991,Balbus1998}, and extensive research on MRI-driven turbulence has been carried out. 
However, even 30 years after the first discovery of MRI-driven accretion turbulence, many unanswered questions remain about the nature of magnetorotational turbulence.

In general, turbulence is characterized by three scales: the energy injection range, inertial range, and dissipation range. 
Since nonlinear effects are dominant in the inertial range, an essential aspect of direct numerical simulation is how to widely resolve the inertial range.
Many theoretical models for the inertial range of magnetohydrodynamics (MHD) turbulence have been proposed~\citep[e.g.,][see also \citealt{Schekochihin2020} for a recent review]{Iroshnikov1963,Kraichnan1965,Goldreich1995,Boldyrev2006,Mallet2017,Loureiro2017}, and direct numerical simulations with free decay or external forcing have confirmed these models.
However, these theoretical models of the inertial range have not been confirmed numerically for MRI-driven MHD turbulence. 

The simulation of magnetorotational turbulence can be divided into two types: global simulations, which solve for the entire disk~\citep[e.g., ][and numerous other recent studies motivated by the Event Horizon Telescope]{Machida2000,Hawley2000,Tchekhovskoy2011,Suzuki2014,Ressler2015,Sadowski2017,Chael2018}, and local shearing box simulations, which clip a part out of the disk for finer resolution~\citep[e.g., ][]{Hawley1995,Sano2004,Sharma2006,Lesur2007,Bodo2008,Hoshino2015,Kunz2016,Zhdankin2017}. 
However, numerical resolution in recent studies is insufficient to reach the inertial range in MRI-driven turbulence even with the shearing box approach\footnote{It has been found that the non-local energy transfer makes it difficult to reach the inertial range in MRI-driven turbulence~\citep{Lesur2011}.}. 
We need a numerical code with high-order accuracy and high-parallel computing performance to resolve the inertial range in MRI-driven turbulence\footnote{Note that a high-order MHD solver using a compact finite difference scheme was developed recently, and the code demonstrated a very narrow dissipation range in MRI-driven shearing-box turbulence~\citep{Hirai2018}.}.
In this study, we develop a code using a pseudospectral method~\citep{Orszag1969}, a highly accurate scheme commonly used in MHD turbulence simulations. 
In the pseudospectral method, a Fourier transform is performed in the spatial direction, and the time evolution is then solved for a finite number of Fourier coefficients. 
The pseudospectral method converges faster than any finite difference scheme for smooth functions~\citep[infinite-order accuracy or spectral accuracy; ][]{Hussaini1987}, resulting in minimal numerical dissipation.
In addition, hyper-viscosity and hyper-resistivity proportional to $\nbl^{2n}$ can be easily used, as algebraic operations replace spatial differentiation, effectively broadening the inertial range.
Moreover, the divergence-free condition of the magnetic field, which is often difficult to implement in finite difference MHD simulations, can be easily satisfied algebraically. 
A negative aspect of the pseudospectral method is that the boundary conditions are restricted because the field is expanded by global orthogonal basis functions. 
In addition, numerical oscillations due to the Gibbs phenomenon occur when discontinuous structures, such as shocks, are present. 
For the former concern, the pseudospectral method can be used for local simulations of accretion disks because periodic boundary conditions can be imposed by transforming the computational domain to shearing coordinates~\citep{Rogallo1981,Umurhan2004}, as described later in this paper.
Regarding shocks, even though MRI-driven turbulence is predominantly subsonic, spiral density waves lead to shock formation~\citep{Heinemann2009}; whether these shocks damage the simulations of our code requires investigation. 
At present, several simulation codes can perform local simulations of accretion disks, but only SNOOPY~\citep{Lesur2007}\footnote{https://ipag.osug.fr/~lesurg/snoopy.html} uses the pseudospectral method.
SNOOPY has been used to solve a variety of plasma turbulence problems~\citep[e.g.,][]{Squire2015,St-Onge2020,Squire2020,Hosking2020,Perrone2021a,Perrone2021b} as well as MRI-driven turbulence using incompressible MHD~\citep{Lesur2011,Kunz2013,Walker2016,Walker2017,Kempski2019,Zhdankin2017}.

In the pseudospectral method, it is necessary to perform fast Fourier transforms (FFTs) and inverse FFTs at each step to evaluate the nonlinear terms. 
However, the three-dimensional FFT is a challenging operation to perform in massively parallel computations because of its algorithmic complexity\footnote{When a strong mean magnetic field exists, a finite difference method may be used in that direction~\citep[e.g.,][]{Numata2010,Chen2011,Loureiro2016,Kawazura2018}. Although this avoids three-dimensional FFTs, it creates difficulty in handling the linear terms implicitly. In addition, when there is no mean magnetic field, there is no reason to treat only one dimension as a special case.}.
Usually, the parallel three-dimensional FFT is performed by the combination of serial FFTs and transposes. 
Since the grids are not parallelized in the directions in which the FFT is performed, the grids are decomposed either in one dimension (slab decomposition) or two dimensions (pencil decomposition).
In the case of slab decomposition, one can use up to $N$ Message Passing Interface (MPI) processes for $N^3$ grids. 
Although a hybrid method with OpenMP increases the number of available cores to some extent~\citep{Mininni2011}, the increase is limited to only a factor of a few in many cases. 
The SNOOPY code adopts the slab decomposition approach.
Pencil decomposition, on the other hand, in principle allows up to $N^2$ MPI processes to be used for $N^3$ grids, enabling much larger parallel computations. 
One of the most popular current FFT libraries using a pencil decomposition approach is P3DFFT~\citep{Pekurovsky2012}\footnote{https://p3dfft.readthedocs.io}.
This study reports the development of an open-source code, \calliope\footnote{https://github.com/ykawazura/calliope}, which solves MRI-driven turbulence in shearing coordinates using a pseudospectral method with the P3DFFT library.
It is expected that this code will be able to compute MRI-driven turbulence at higher resolutions than those previously used. 
To the best of our knowledge, no other pseudospectral code solves MRI-driven turbulence using pencil decomposition. 
Thus, we believe that the release of \calliope~as an open-source code will benefit the astrophysical community. 
Furthermore, \calliope~can also be applied to the study of other kinds of three-dimensional MHD turbulence.

This paper is organized as follows.
In Section~\ref{s:model}, we discuss the models that \calliope~can solve. 
Next, in Section~\ref{s:scheme}, we describe the numerical algorithm used by \calliope. 
The results of linear and nonlinear numerical tests are then presented in Section~\ref{s:tests}.
Finally, Section~\ref{s:summary} summarizes the study’s conclusions.

\section{Model} \label{s:model}
In this section, we describe the models that \calliope~can solve: isothermal compressible MHD, incompressible MHD, and rotational reduced MHD~\citep[RRMHD; ][]{Kawazura2021}. 
First, we consider a system of Cartesian coordinates that co-rotates with the disk at a distance $r_0$ from the center of the disk with an angular velocity $\Omega\zhat$. 
In this system, the coordinate axes $(x, y, z)$ are taken in the radial, azimuthal, and vertical directions, respectively. 
The set of equations for isothermal compressible MHD is:
\begin{subequations}
\begin{align}
  &\pp{\rho}{t} + \bm{u}_0\cdot\nbl\rho = -\Div\bm{M},
  \label{e:MHD_COMP_ISOTH rho}\\
  &\pp{\bm{M}}{t} + \bm{u}_0\cdot\nbl\bm{M} = -\Div\lf[\rho\bm{u}\bm{u} - \f{\bm{B}\bm{B}}{4\pi} + 
\lf(c_\rmS^2\,\rho + \f{B^2}{8\pi}\ri)\mathbb{I}\ri] \nonumber\\
	&\hspace{12em} - 2\Omega\unit{z}\times\bm{M} - \bm{M}\cdot\nbl\bm{u}_0,
  \label{e:MHD_COMP_ISOTH M}\\
  &\pp{\bm{B}}{t} + \bm{u}_0\cdot\nbl\bm{B} = -\Curl\lf(\bm{B}\times\bm{u}\ri) + \bm{B}\cdot\nbl\bm{u}_0,
  \label{e:MHD_COMP_ISOTH B}\\
  &\Div\bm{B} = 0,
  \label{e:MHD_COMP_ISOTH div B}
\end{align}
\end{subequations}
where $\bm{u}_0 = -q\Omega\unit{y}$ is the background shear flow, $q = -(\rmd \ln \Omega/\rmd \ln r)_{r = r_0}$ is the shear rate, $\rho$ is the density, $\bm{u}$ is the velocity, $\bm{M} \equiv \rho\bm{u}$ is the momentum density, $\bm{B}$ is the magnetic field, $c_\rmS$ is the sound speed (constant), and $\mathbb{I}$ is the unit tensor. 
In the following, we consider only the case of Keplerian rotation, i.e., $q = 3/2$. 

Next, the set of equations for the incompressible MHD is:
\begin{subequations}
\begin{align}
  &\pp{\bm{u}}{t} + \bm{u}_0\cdot\nbl\bm{u} = -\Div\lf(\bm{u}\bm{u} - \f{\bm{B}\bm{B}}{4\pi\rho} + P\,\mathbb{I}\ri) - 2\Omega\unit{z}\times\bm{u} - \bm{u}\cdot\nbl\bm{u}_0,
  \label{e:MHD_INCOMP u}\\
  &\pp{\bm{B}}{t} + \bm{u}_0\cdot\nbl\bm{B} = -\Curl\lf(\bm{B}\times\bm{u}\ri) + \bm{B}\cdot\nbl\bm{u}_0,
  \label{e:MHD_INCOMP B}\\
  &\Div\bm{u} = 0, \; \Div\bm{B} = 0.
  \label{e:MHD_INCOMP div u div B}
\end{align}
\end{subequations}
where $\rho$ is the constant density and $P$ is the thermal pressure to be determined by $\Div \bm{u} = 0$.

In the shearing box, we impose periodic boundaries in the $y$- and $z$-directions and a shearing boundary condition $f(0, y, z) = f(L_x, y - q\Omega L_x t, z)$ in the $x$-direction~\citep{Hawley1995}, where $L_x$ is the box size in the $x$-direction. 
In order to use the pseudospectral method, the $x$-direction must also be periodic, therefore, we perform the shearing coordinate transformation $y \mapsto y - q\Omega t x$~\citep{Rogallo1981,Umurhan2004}. 
By this transformation, $x$-direction becomes periodic, and the second term on the left-hand side of equations \eqref{e:MHD_COMP_ISOTH rho}-\eqref{e:MHD_COMP_ISOTH B} and \eqref{e:MHD_INCOMP u}-\eqref{e:MHD_INCOMP B} disappears; instead, the wavenumber in the $x$-direction evolves in time, as described in the next section.

Next, we show the RRMHD equations. 
Unlike the other models presented above, we assume the presence of a background magnetic field $\bm{B}_0$ that is constant in time and space and tilted at an angle $\theta$ with respect to the equatorial plane of the accretion disk. 
We define $x'$ as the radial direction, $z'$ as the $\bm{B}_0$ direction, and $y'$ as the direction perpendicular to $x'$ and $z'$. 
We then employ the Reduced MHD (RMHD) approximation; namely, we make assumptions regarding the wavenumber anisotropy ($k_\|/k_\+ \ll 1$) and small amplitude fluctuations ($\delta\bm{B}/B_0 \sim \bm{u}/v_\rmA \ll 1$), where $k_\|$ ($k_\+$) is the parallel (perpendicular) wavenumber component to $\bm{B}_0$, $\delta\bm{B}$ represents the magnetic field fluctuations, and $v_\rmA = B_0/\sqrt{4\pi\rho_0}$ is the Alfv\'en speed. 
We further assume that $\bm{B}_0$ is almost azimuthal, i.e., $\sin\theta \ll 1$.
Under these assumptions, the set of equations for RRMHD is:
\begin{subequations}
\begin{align}
  &\lf(\pp{}{t} + \bm{u}_\+\cdot\nbl'_\+\ri)\Psi = v_\rmA\pp{\Phi}{z'},
  \label{e:RRMHD Psi}\\
  &\lf(\pp{}{t} + \bm{u}_\+\cdot\nbl'_\+\ri){\nbl'_\+}^2\Phi = v_\rmA \lf(\pp{}{z'} + \f{\delta\bm{B}_\+}{B_0}\cdot\nbl'_\+\ri){\nbl'_\+}^2\Psi \nonumber\\ 
  &\hspace{15em} - 2\Omega\sin\theta\,\pp{u_\|}{y'},
  \label{e:RRMHD Phi}\\
  &\lf(\pp{}{t} + \bm{u}_\+\cdot\nbl'_\+\ri)u_\| = v_\rmA^2\lf(\pp{}{z'} + \f{\delta\bm{B}_\+}{B_0}\cdot\nbl'_\+\ri)\f{\delta B_\|}{B_0} \nonumber \\
  &\hspace{13em} + (2 - q)\Omega \sin\theta\,\pp{\Phi}{y'},
  \label{e:RRMHD u}\\
  &\lf(\pp{}{t} + \bm{u}_\+\cdot\nbl'_\+\ri)\lf(  1 + \f{v_\rmA^2}{c_\rmS^2} \ri)\f{\delta B_\|}{B_0} = \lf(\pp{}{z'} + \f{\delta\bm{B}_\+}{B_0}\cdot\nbl'_\+\ri) u_\| \nonumber \\
  &\hspace{15em} + \f{q\Omega \sin\theta}{v_\rmA}\,\pp{\Psi}{y'},
  \label{e:RRMHD B}
\end{align}
\end{subequations}
where $\Phi$ and $\Psi$ are the stream and flux functions, respectively, defined by $\bm{u}_\+ = \zhat'\times\nbl'_\+\Phi$ and $\delta\bm{B}_\+ = \sqrt{4\pi\rho}\zhat'\times\nbl'_\+\Psi$. 
When the angular velocity is zero (i.e., $\Omega = 0$), the RRMHD becomes RMHD, where \eqref{e:RRMHD Psi}-\eqref{e:RRMHD Phi} and \eqref{e:RRMHD u}-\eqref{e:RRMHD B} are decoupled, and $u_\|$ and $\delta B_\|$ are passive with respect to $\Phi$ and $\Psi$~\citep{Schekochihin2009}. 
In RRMHD, the effect of the background shear flow disappears due to the assumptions that $k_\|/k_\+ \ll 1$ and $\sin\theta \ll 1$, so the periodic boundary condition can be imposed without transforming to shearing coordinates.

\section{Numerical scheme} \label{s:scheme}
Since all the models described above do not include dissipation, meaning that energy accumulates at the grid-scale due to the turbulent cascade, hyper-dissipation terms should be added to the right-hand side of each model when performing simulations with \calliope. 
For the isothermal compressible MHD and incompressible MHD cases, the hyper-dissipation is proportional to $\nbl^{2n}$, where $n$ is an integer greater than or equal to unity. 
For the RRMHD case, the perpendicular hyper-dissipation proportional to $\nbl_\+^{2n}$ and the parallel hyper-dissipation proportional to $(\p/\p z)^{2n}$ can be set independently.

Let $\bm{U}$ be the set of the field variables [e.g., $\bm{U} = (\rho,\, \bm{M},\, \bm{B})$ for isothermal compressible MHD]. The models solved by \calliope~can then be expressed as:
\begin{equation}
  \dd{\bm{U}_\bm{k}}{t} = \lf(\calN[\bm{U}]\ri)_\bm{k} + \calL[\bm{U}_\bm{k}] + \calD[\bm{U}_\bm{k}]
  \label{e:dUdt}
\end{equation}
where the $\bm{U}_\bm{k}$ denote the Fourier coefficients of $\bm{U}$, $\lf(\calN[\bm{U}]\ri)_\bm{k}$ denote the Fourier coefficients of the nonlinear terms, $\calL[\bm{U}_k]$ is the linear term originating from the rotation, and $\calD[\bm{U}_k]$ is the hyper-dissipation term, respectively. 
In \calliope, $\calD$ is treated implicitly, and $\calN$ and $\calL$ are treated explicitly. 
The time evolution is solved using the third-order Gear method~\citep{Karniadakis1991}:
\begin{multline}
  \f{1}{\Delta t}\lf(\f{11}{6}\bm{U}_\bm{k}^{n+1} - 3\bm{U}_\bm{k}^{n} + \f{3}{2}\bm{U}_\bm{k}^{n-1} - \f{1}{3}\bm{U}_\bm{k}^{n-2}\ri) \\= 3\lf(\calN[\bm{U}]_\bm{k}^{n} + \calL[\bm{U}_\bm{k}^{n}]\ri) - 3\lf(\calN[\bm{U}]_\bm{k}^{n-1} + \calL[\bm{U}_\bm{k}^{n-1}]\ri) \\
  + \lf(\calN[\bm{U}]_\bm{k}^{n-2} + \calL[\bm{U}_\bm{k}^{n-2}]\ri) + \calD[\bm{U}_\bm{k}^{n+1}],
  \label{e:gear}
\end{multline}
where the superscript $n$ denote the value at the $n$-th timestep.
To remove the aliasing error of the nonlinear term, we adopt a 2/3-rule utilizing the pruned-FFT feature of P3DFFT.

\begin{figure}
  \begin{center}
    \includegraphics*[width=0.47\textwidth]{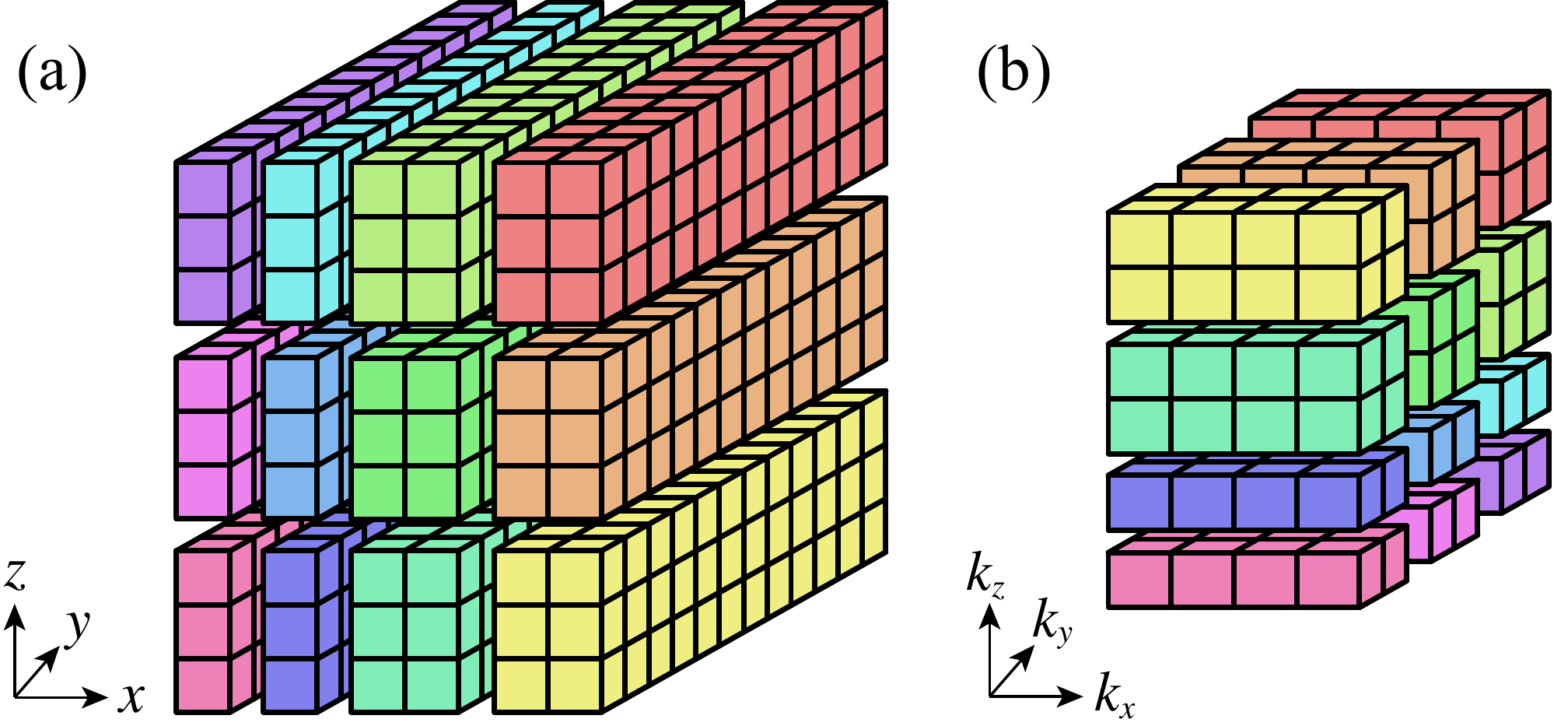}
  \end{center}
  \caption{Schematic of pencil decomposition in \calliope. (a) and (b) correspond to the physical and Fourier space grids, respectively.}
  \label{f:pencil decomposition}
\end{figure}

We now describe the array layout of the field variables. 
\calliope~solves the time evolution of the fields in the Fourier space while the physical space is used only to evaluate the nonlinear terms. 
In pencil decomposition, one dimension of the three-dimensional array is not distributed between MPI processes, i.e., the array is localized in that direction.
As shown in Fig.~\ref{f:pencil decomposition}, the $y$-direction in physical space and the $k_x$-direction in Fourier space are localized. 
This choice is to avoid unnecessary transposition or inter-process communication during the remapping process, as described below. 
Since \calliope~uses the stride-1 data structure of P3DFFT, the array layout in each process is:
\begin{equation*}
  \begin{array}{c}
    \text{Physical space} \\
    \lf(N^{(l)}_y,\, N^{(l)}_z/M_1,\, N^{(l)}_x/M_2\ri)
  \end{array} 
  \;\;
  \Longleftrightarrow
  \;\
  \begin{array}{c}
    \text{Fourier space} \\
     \lf(N^{(k)}_x,\, N^{(k)}_z/M_2,\, N^{(k)}_y/M_1\ri)
  \end{array} 
  \label{e:array layout}
\end{equation*}
where $N^{(l)}_x, N^{(l)}_y$, and $N^{(l)}_z$ are the numbers of grids in physical space, and $N^{(k)}_x = 2N^{(l)}_x/3, N^{(k)}_y = N^{(l)}_y/3 + 1$, and $N^{(k)}_z = 2N^{(l)}_z/3$ are the numbers of grids in Fourier space. 
In addition, $M_1 M_2$ represents the total number of processes.

Next, the remapping method is described. 
In \calliope, the periodic remapping method~\citep{Rogallo1981,Umurhan2004} is used for isothermal compressible MHD and incompressible MHD. 
This method is also used in the SNOOPY code. 
In shearing coordinates, the wavenumber in the $x$-direction evolves in time according to $k_x(t) = k_x + q\Omega t k_y$. 
To prevent $k_x(t)$ from growing limitlessly, remapping is performed every $T = L_y/q\Omega L_x$. 
The fields that are periodic in the $x$-direction in the non-shearing coordinate system at $t = 0$ become periodic again in the $x$-direction in the non-shearing coordinate system at $t = T$.
Thus, we can rearrange the field such that:
\begin{equation}
  f(k_x,\, k_y,\, k_z) \mapsto f(k_x + k_y L_x/L_y,\, k_y,\, k_z).
\end{equation}
Simultaneously, we reset $k_x(T)$ to $k_x(0)$. 
With this rearrangement, the data outside the computational domain are discarded, and the portion newly allocated to the computational domain is initialized to zero.
Thus, the time evolution of the model becomes somewhat choppy before and after remapping\footnote{To avoid this, continuous remapping~\citep{Lithwick2007} and the phase-shifting Fourier Transform~\citep{Brucker2007} have been proposed, but implementing these into \calliope~is a future task.}. 
Since the array is rearranged in the $k_x$ direction upon remapping, and \calliope~localizes the array in the $k_x$ direction, data does not need to be transferred between processors when remapping.

Note, finally, that \calliope~is currently limited to periodic boundary conditions in all three directions. 
P3DFFT can support a Chebyshev transform in one direction allowing non-periodic boundaries while the other two directions are Fourier transformed. 
This set of transforms is useful to solve systems in a spherical shell. 
Implementation of a Chebyshev transform in \calliope~should be conducted in the future.

\section{Tests} \label{s:tests}
In this section, we show the results of linear and nonlinear tests and demonstrate the parallel performance of \calliope.

\subsection{Linear wave propagation in isothermal MHD} \label{ss:linear waves}
\begin{figure*}
  \begin{center}
    \includegraphics*[width=1.0\textwidth]{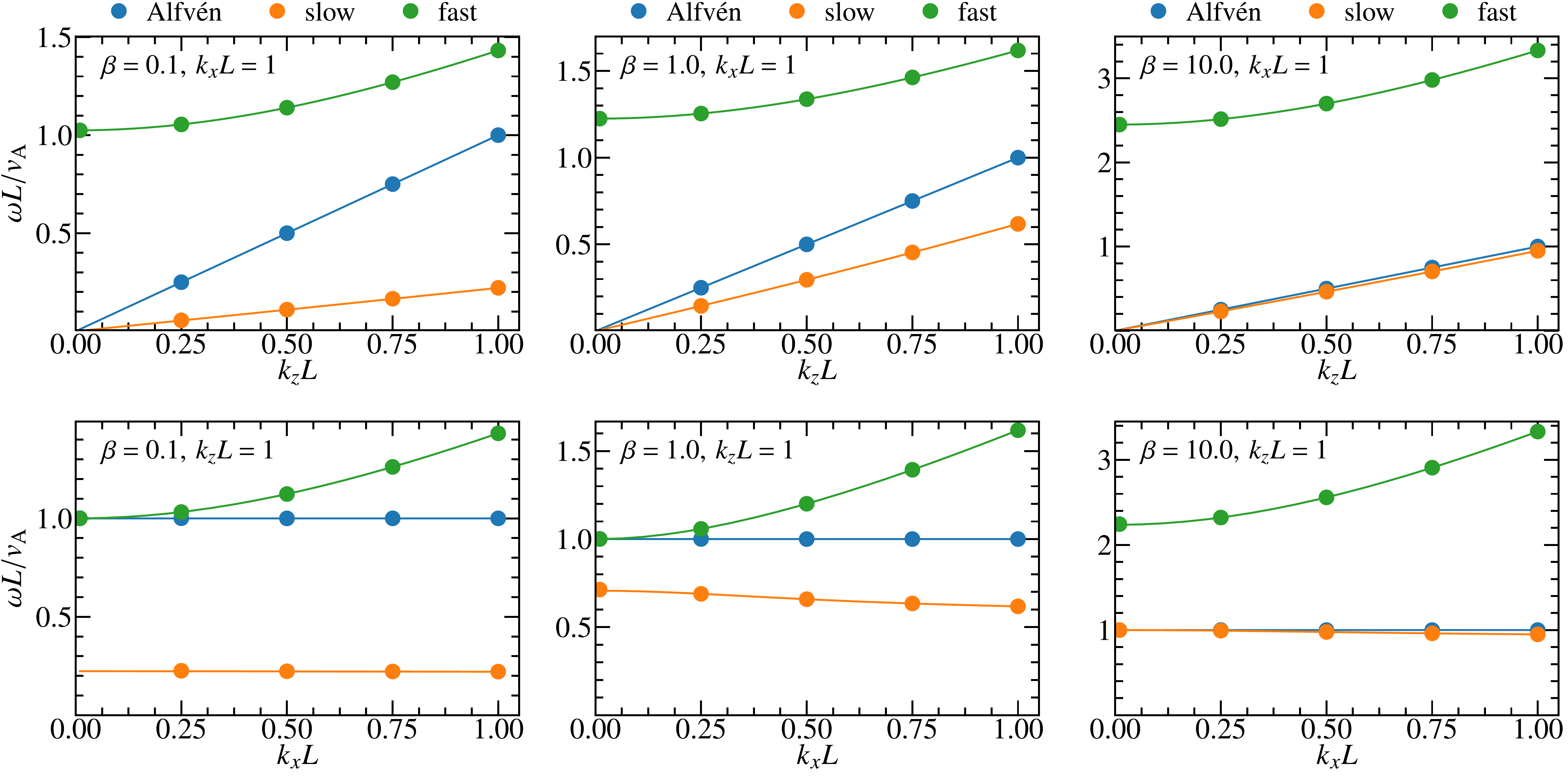}
  \end{center}
  \caption{Linear wave propagation test of isothermal compressible MHD. Markers are numerically obtained frequencies, and solid lines are theoretical values.}
  \label{f:linear waves}
\end{figure*}

As a first relatively simple test, we calculate linear wave propagation in isothermal compressible MHD. 
We set the uniform background magnetic field $(0, 0, B_0)$, the uniform background density $\rho_0$, and the wavenumber vector $(k_x, 0, k_z)$. 
The initial perturbations of the fields are set to the eigenfunctions of the Alfv\'en, slow, and fast modes, as shown below~\citep{Goedbloed2004}
\begin{itemize}
  \item Alfv\'en mode
    \begin{align*}
      \f{B_y}{B_0} = -\f{u_y}{v_\rmA},\;\; u_x = u_z = B_x = B_z = \rho = 0
    \end{align*}
  \item Slow and fast modes
    \begin{align*}
      & \f{\rho}{\rho_0} = \lf( \f{\alpha_{\rms, \rmf}\omega_{\rms, \rmf}}{k_x v_\rmA} \ri)\lf( \f{v_\rmA}{c_\rmS} \ri)^2\f{u_x}{v_\rmA}, \;\; \f{B_z}{B_0} = \f{k_x u_x}{\omega_{\rms, \rmf}}, \;\;  \\
      & u_z = \f{\alpha_{\rms, \rmf}k_\|}{k_x}u_x,\;\; u_y = B_y = 0,
    \end{align*}
\end{itemize}
where $\omega_{\rms,\rmf}$ is the frequency of the slow and fast modes:
\begin{equation*}
  \omega_{\rms,\rmf} = k\sqrt{\f{1}{2}\lf[v_\rmA^2 + c_\rmS^2 \pm \sqrt{\lf(v_\rmA^2 + c_\rmS^2\ri)^2 - 4\lf(k_z^2/k^2\ri) v_\rmA^2 c_\rmS^2}\ri]}  
\end{equation*}
and $\alpha_{\rms,\rmf} = 1 - k^2 v_\rmA^2/\omega_{\rms,\rmf}^2$.
The subscripts s and f denote the slow and fast modes, corresponding to the minus and plus signs in the above equation, respectively. 
We initialize a mode with one of these eigenfunctions and compute the time evolution to obtain the frequency. 
We test cases where the value of $\beta \equiv 8\pi \rho_0 c_\rmS^2/B_0^2$ is 0.1, 1, and 10 by fixing $k_x L = 1$ and varying $k_z$, and by fixing $k_z L = 1$ and varying $k_x$. 
Figure~\ref{f:linear waves} shows the results of these tests. 
In all cases, the numerically obtained frequencies accurately reproduce the theoretical values.

\subsection{Axisymmetric linear MRI in incompressible MHD} \label{ss:linear MRI}
\begin{figure}
  \begin{center}
    \includegraphics*[width=0.45\textwidth]{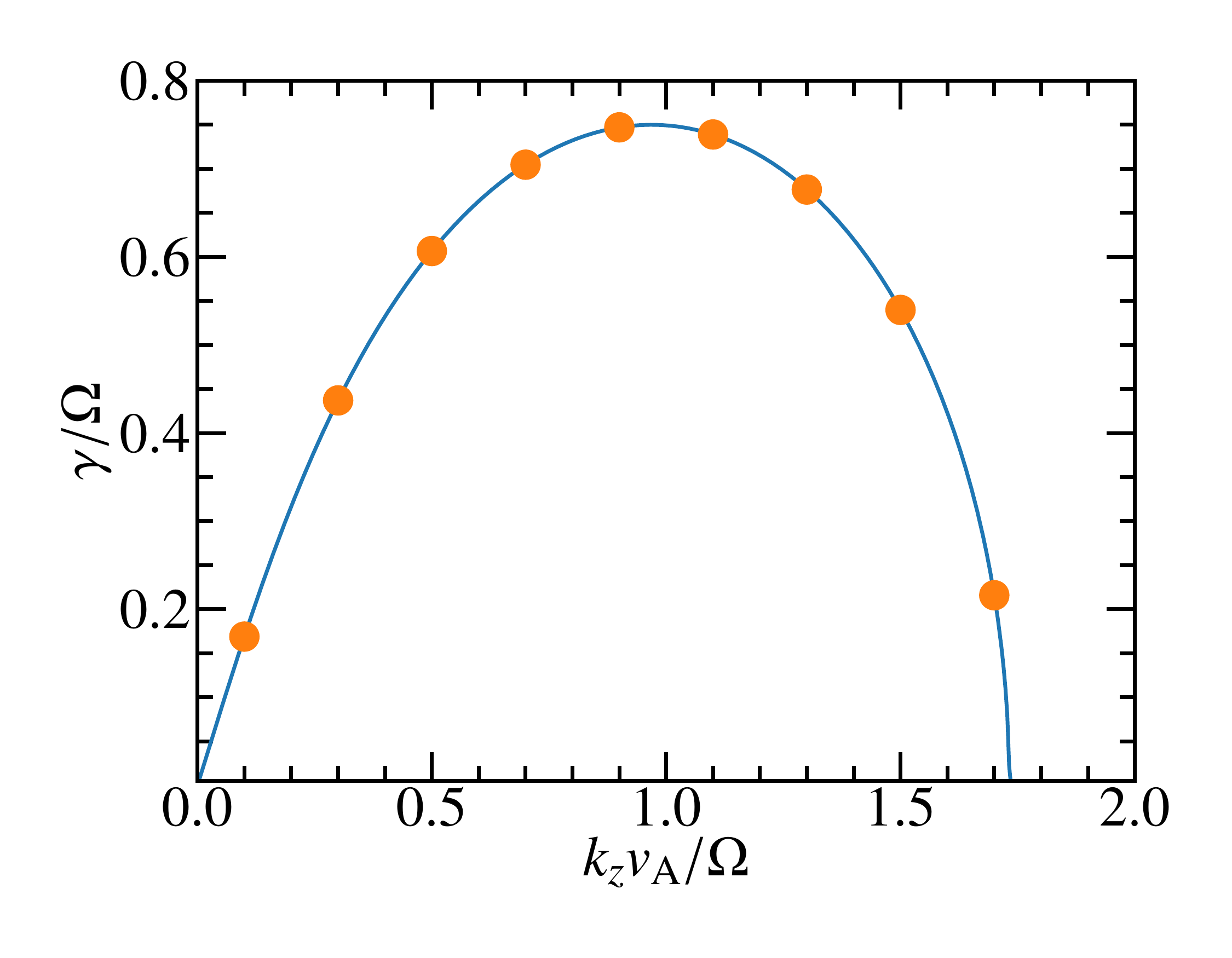}
  \end{center}
  \caption{Axisymmetric linear MRI test of incompressible MHD. Markers indicate the numerically obtained growth rates, and the solid line represents the theoretical values.}
  \label{f:linear MRI}
\end{figure}

Next, we present a test of an axisymmetric ($k_y = 0$) linear MRI in incompressible MHD. 
The initial magnetic field is set to $(0, 0, B_0)$. 
The wavenumber is considered only in the $z$-direction. 
In this case, the eigenfunctions of the MRI are given by:
\begin{align*}
  &B_y = \lf[ \f{\omega^2 - (k_z v_\rmA)^2 + 2q\Omega^2}{2\rmi\Omega\omega} \ri]B_x, \;\; B_z = 0,\\
  &u_x = -\f{\omega}{k_z}\f{B_x}{B_0}, \;\; u_y = -\f{1}{k_z}\lf( \omega\f{B_y}{B_0} + \rmi q\Omega\f{B_x}{B_0} \ri), \;\; u_z = 0,
\end{align*}
where
\begin{equation}
  \omega = \rmi\gamma = \rmi\sqrt{\sqrt{4\lf( k_z v_\rmA \ri)^2\Omega^2 + (2 - q)\Omega^4} - \lf( k_z v_\rmA \ri)^2 - (2 - q)\Omega^2},
\end{equation}
and $\gamma$ is the growth rate of the MRI.
One mode is initialized with this eigenfunction, and the time evolution is calculated to obtain the growth rate. 
As shown in Fig.~\ref{f:linear MRI}, the numerically computed growth rates accurately reproduce the theoretical values.

\subsection{Two-dimensional Orszag--Tang vortex problem of RMHD} \label{ss:OT2}
\begin{figure}
  \begin{center}
    \includegraphics*[width=0.4\textwidth]{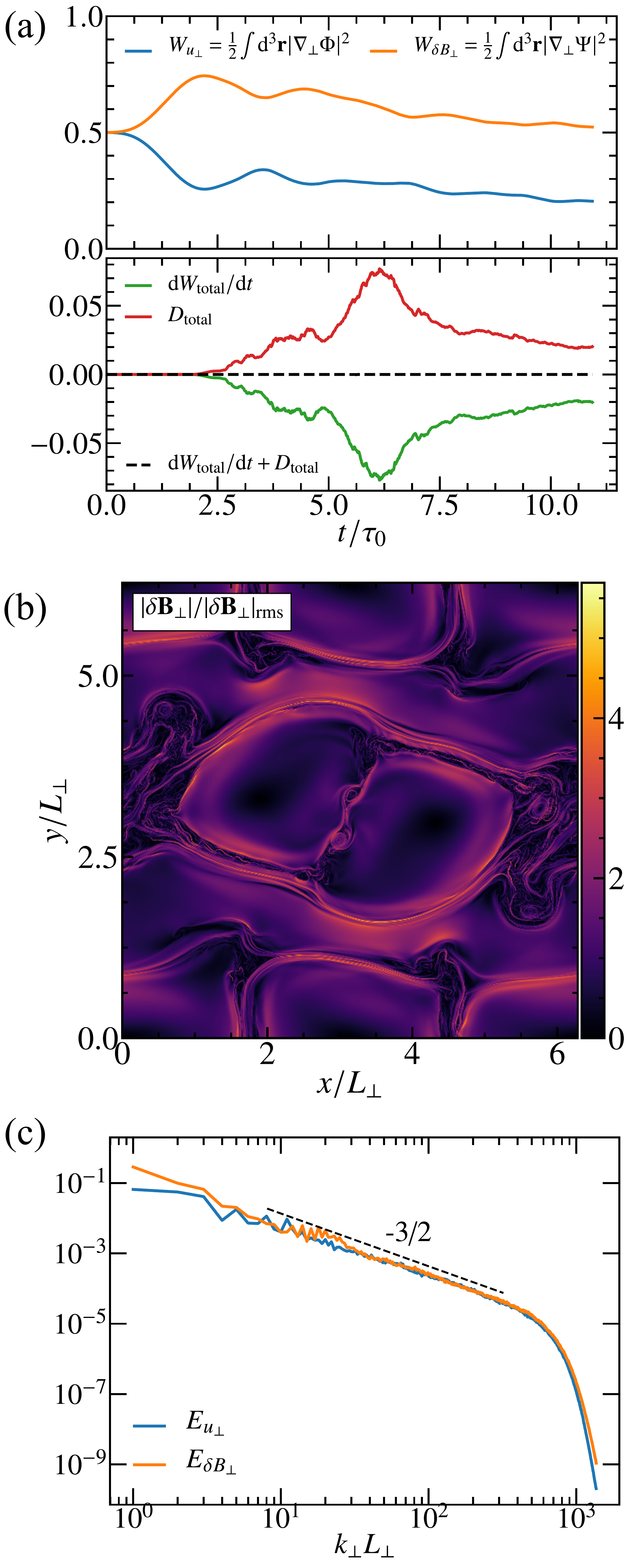}
  \end{center}
  \caption{Results of the two-dimensional Orszag--Tang vortex problem using RMHD. (a) Time evolution of magnetic and kinetic energy, and power balance (upper and lower panel, respectively), (b) spatial profile of the magnetic field strength at $t \simeq 6\tau_0$, and (c) energy spectra of velocity and magnetic fields. The dashed line shows the $k_\+^{-3/2}$ power law.}
  \label{f:OT2}
\end{figure}

In the following section, we present the tests of nonlinear simulations. 
First, we compute the two-dimensional Orszag-–Tang problem using the RMHD model~\citep{Orszag1979}. 
We initialize $\Phi$ and $\Psi$ as follows:
\begin{eqnarray}
  \Phi &=& - u_0\lf( \f{L_\+}{2\pi} \ri) \lf[ \cos\lf( \f{2\pi x}{L_\+} \ri) + \cos\lf( \f{2\pi y}{L_\+} \ri) \ri], \nonumber\\
  \Psi &=& - u_0\lf( \f{L_\+}{2\pi} \ri) \lf[ \f{1}{2}\cos\lf( \f{4\pi x}{L_\+} \ri) + \cos\lf( \f{2\pi y}{L_\+} \ri) \ri],
\end{eqnarray}
where $u_0$ is the initial speed, and $L_\+$ is the size of $x$ and $y$ directions.
Hyperdissipation proportional to $\nbl_\+^8$ is used to terminate the turbulent cascade. 
Figure~\ref{f:OT2} shows the results of the test with $(n^{(l)}_x,\, n^{(l)}_y) = (2048,\, 2048)$. 
In Fig.~\ref{f:OT2}-(a), we plot each term of the time evolution of free energy:
\begin{equation}
  W_\mr{tot} = W_{u_\+} + W_{\delta B_\+} = \f{1}{2}\int\rmd^3\bm{r}\,\lf( |\nbl_\+\Phi|^2 + |\nbl_\+\Psi|^2 \ri),
\end{equation}
and power balance:
\begin{equation}
  \dd{W_\mr{tot}}{t} = -D_\mr{tot},
\end{equation}
where $-D_\mr{tot}$ is the sum of the dissipation due to the hyper-viscosity and hyper-resistivity terms. 
At early times, the magnetic field energy increases and the kinetic energy decreases, with the magnetic field energy reaching a peak at $t \simeq 2\tau_0$ where $\tau_0 = L_\+/u_0$. 
This behavior is consistent with the results of a simulation by~\citet{Parashar2009}. 
When $t \lesssim 2\tau_0$, only energy exchange occurs between the magnetic field and flow, with no energy dissipation [lower panel in Fig.~\ref{f:OT2}-(a)]. 
The cascade then reaches the grid scale at $t \simeq 2\tau_0$, and the energy dissipation starts to increase.
The magnetic field profile at $t \simeq 6\tau_0$ is shown in Fig.~\ref{f:OT2}-(b), where vortices of various sizes ranging from box scale to grid scale are present, indicating that the flow is turbulent. 
Fig.~\ref{f:OT2}-(c) shows the kinetic and magnetic spectra at $t \simeq 6\tau_0$. 
Both spectra follow $k_\+^{-3/2}$ power law, which is consistent with observations from other studies of two-dimensional freely decaying MHD turbulence~\citep[e.g.,][]{Biskamp2001}. 
The spectra also show that the dissipation is suppressed until the cascade approaches the grid scale, resulting in a wide inertial range.

\subsection{Three-dimensional nonlinear MRI turbulence in isothermal MHD} \label{ss:nonlinear MRI}
In this section, we present a test of three-dimensional MRI-driven turbulence in isothermal compressible MHD. 
This is the first study where compressible MRI turbulence has been simulated using the pseudospectral method, to the best of our knowledge. 
The sound speed $c_\rmS$, the angular velocity $\Omega$, initial density $\rho_0$, and uniform initial magnetic field $(0, 0, B_0)$ are set such that the condition $\lambda_\mr{MRI} = 0.1H$ is satisfied, where $H = c_\rmS/\Omega$ is the scale height of the accretion disk and $\lambda_\mr{MRI} = 2\pi v_\rmA/\Omega$ is approximately equal to the wavelength of the fastest-growing MRI mode~\citep{Balbus1998}. 
The box size is set to $(2H, 4H, H)$.
The corresponding $\beta$ for this setting is $7.8\times10^3$.
The turbulent cascade is terminated with hyper-dissipation proportional to $\nbl^8$.

\begin{figure}
  \begin{center}
    \includegraphics*[width=0.4\textwidth]{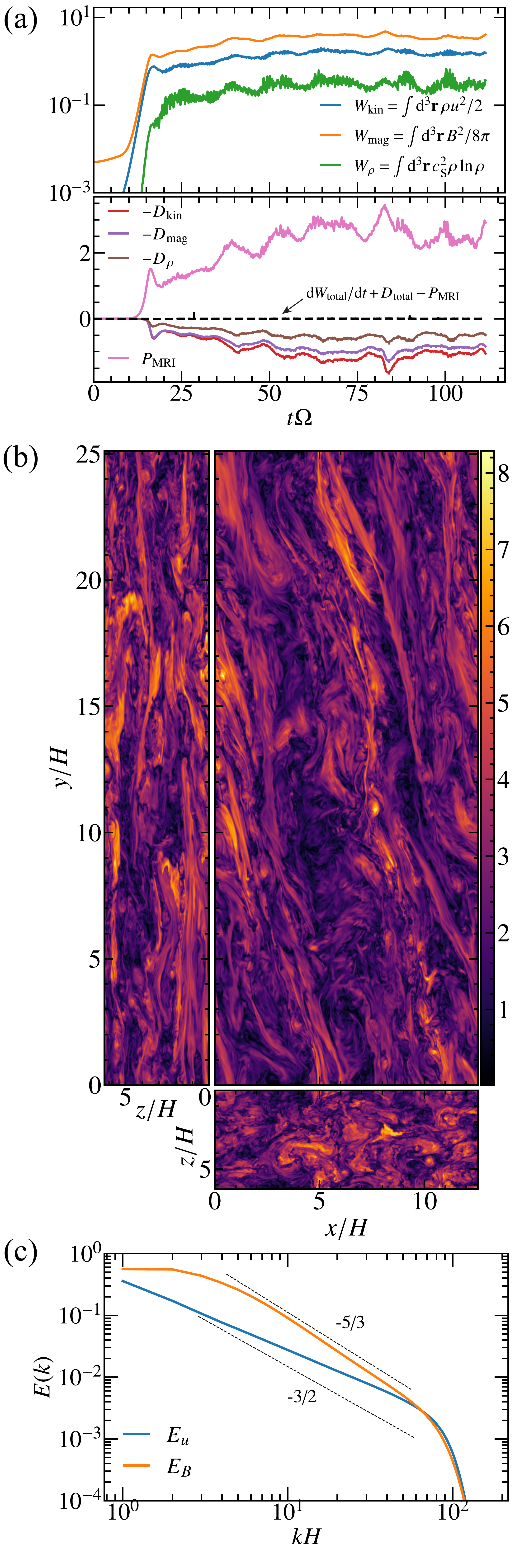}
  \end{center}
  \caption{Results of the three-dimensional MRI-driven turbulence in isothermal compressible MHD. (a) Time evolution of the energy and power balance (upper and lower panels, respectively), (b) spatial profile of the magnetic field strength, and (c) energy spectra of the velocity and magnetic fields. The dashed lines show the $k^{-3/2}$ and $k^{-5/3}$ power laws.}
  \label{f:nolinear MRI}
\end{figure}

Figure~\ref{f:nolinear MRI} shows the results of a test with $(n^{(l)}_x,\, n^{(l)}_y,\, n^{(l)}_z) = (1024,\, 2048,\, 512)$. 
Even though it is a test, the resolution of this simulation is higher than any other previously reported shearing box simulation. 
In Figure~\ref{f:nolinear MRI}-(a), we plot each term of the time evolution of the free energy:
\begin{equation}
  W_\mr{tot} = W_\mr{kin} +  W_\mr{mag} + W_\rho = \int\rmd^3\bm{r}\lf( \f{\rho u^2}{2} + \f{B^2}{8\pi} + c_\rmS^2\,\rho\ln\rho \ri),
\end{equation}
and power balance:
\begin{equation}
  \dd{W_\mr{tot}}{t} = P_\mr{MRI} - D_\mr{total},
\end{equation}
where
\begin{equation}
  P_\mr{MRI} = q\Omega\int\rmd^3\bm{r}\lf( \rho u_x u_y - \f{B_x B_y}{4\pi}\ri)
\end{equation}
is the MRI energy injection rate and $D_\mr{total}$ is the sum of the hyper-dissipation for $\bm{M}$, $\bm{B}$, and $\rho$. 
As shown in Figure~\ref{f:nolinear MRI}-(a), when $t\Omega\lesssim 10$, the MRI grows linearly and then transitions to a nonlinearly saturated state. 
This figure also illustrates that the power balance is precisely maintained at any given time in the simulation.

Figure~\ref{f:nolinear MRI}-(b) shows the spatial profile of magnetic field strength on the $x = 0$, $y = 0$, and $z = 0$ planes. 
The structure shows clear elongation in the $y$-direction, due to the stretching caused by shear flow in the $y$ direction, being consistent with other shearing box simulations.

Finally, Fig.~\ref{f:nolinear MRI}-(c) shows the omni-directional energy spectra of the velocity and the magnetic field. 
The velocity field is shallower than $k^{-3/2}$ and the magnetic field is slightly steeper than $k^{-5/3}$.
These are consistent with the spectra found in previous shearing box simulations~\citep[e.g.,][]{Sun2021}.
In addition, since the cascade is terminated with hyper-dissipation proportional to $\nbl^8$, the dissipation is suppressed until the cascade approaches the grid scale.

\subsection{Parallel Performance} \label{ss:performance}
\begin{figure}
  \begin{center}
    \includegraphics*[width=0.4\textwidth]{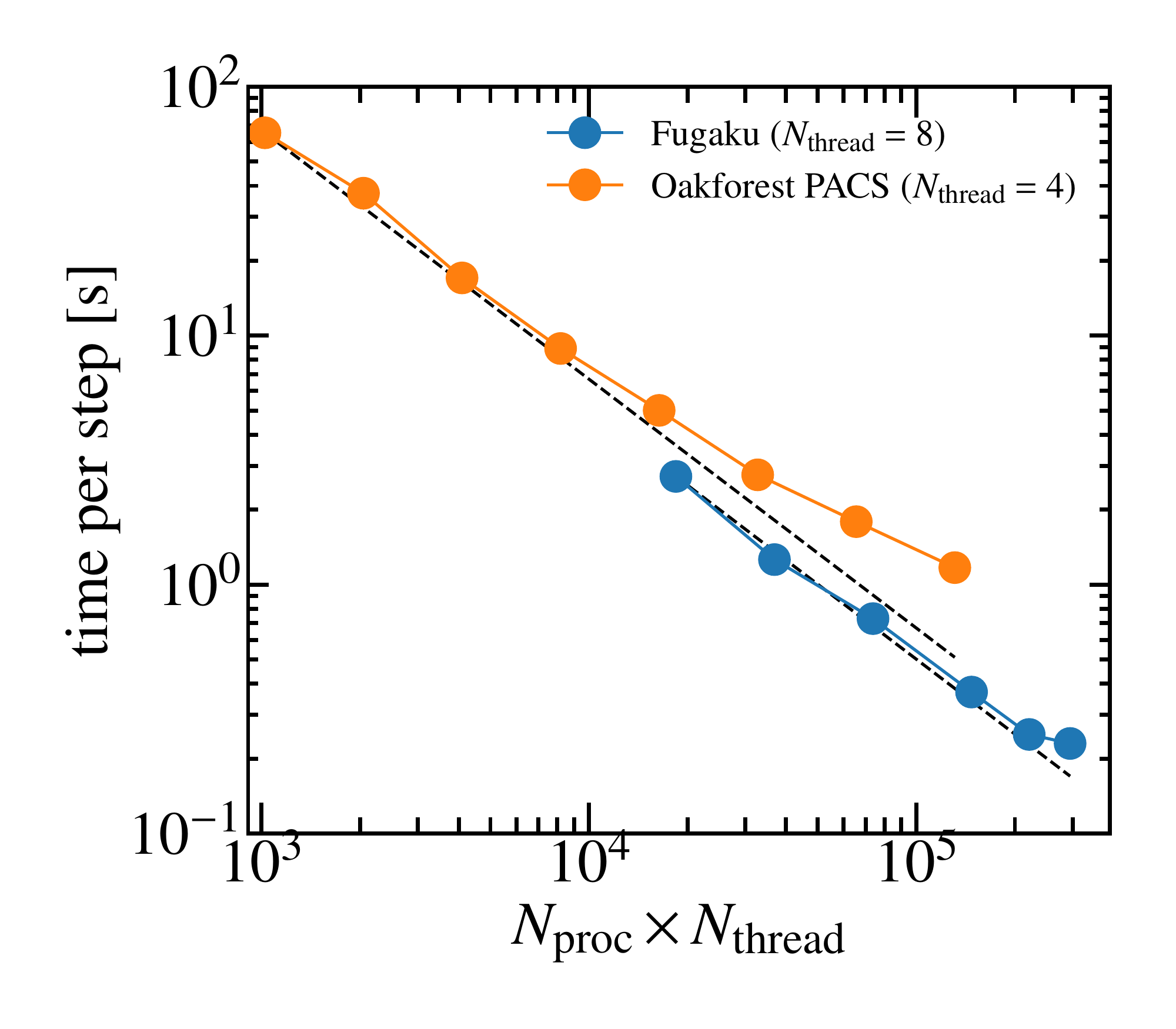}
  \end{center}
  \caption{Strong scaling on Oakforest-PACS and Fugaku for incompressible MHD with $(n^{(l)}_x,\, n^{(l)}_y,\, n^{(l)}_z) = (2048,\, 2048,\, 2048)$. The number of threads is fixed at $N_\mr{thread} = 4$ for Oakforest-PACS and $N_\mr{thread} = 8$ for Fugaku. Dashed lines show the ideal scaling.}
  \label{f:scaling}
\end{figure}

Finally, we describe the parallel performance of \calliope. 
In the pseudospectral method, most of the computation time is consumed evaluating the nonlinear terms using the FFTs and inverse FFTs. 
Therefore, the parallel performance of \calliope~is mostly determined by that of P3DFFT~\footnote{More specifically, the time consumed by P3DFFT is divided into the computation of FFT and communication for a transpose. \citet{Czechowski2012} reported that the latter is dominant over the former.}. 
Here, the parallel performance was measured for nonlinear incompressible MHD simulations with $(n^{(l)}_x,\, n^{(l)}_y,\, n^{(l)}_z) = (2048,\, 2048,\, 2048)$. 
Note that in the case of isothermal compressible MHD, only one forward FFT is added for $\rho$ while the number of inverse FFT remains unchanged, so the scaling is presumably almost the same as that for incompressible MHD. 
The measurements were carried out on Oakforest-PACS\footnote{\href{https://www.cc.u-tokyo.ac.jp/en/supercomputer/ofp/service/}{https://www.cc.u-tokyo.ac.jp/en/supercomputer/ofp/service/}} at the University of Tokyo and Fugaku\footnote{\href{https://www.r-ccs.riken.jp/en/fugaku/}{https://www.r-ccs.riken.jp/en/fugaku/}} (the Japanese flagship supercomputer as of 2021) at RIKEN. 
The highest performance was obtained when the number of threads was $N_\mr{thread} = 4$ on Oakforest-PACS and $N_\mr{thread} = 8$ on Fugaku. 
We fixed the number of threads at these values and measured the execution time upon changing the number of MPI processes, $N_\mr{proc}$. 
Figure~\ref{f:scaling} shows strong scaling, demonstrating the excellent parallel performance of \calliope.
As shown, almost ideal scaling is maintained up to $2\times10^5$ cores at Fugaku.

\begin{figure}
  \begin{center}
    \includegraphics*[width=0.5\textwidth]{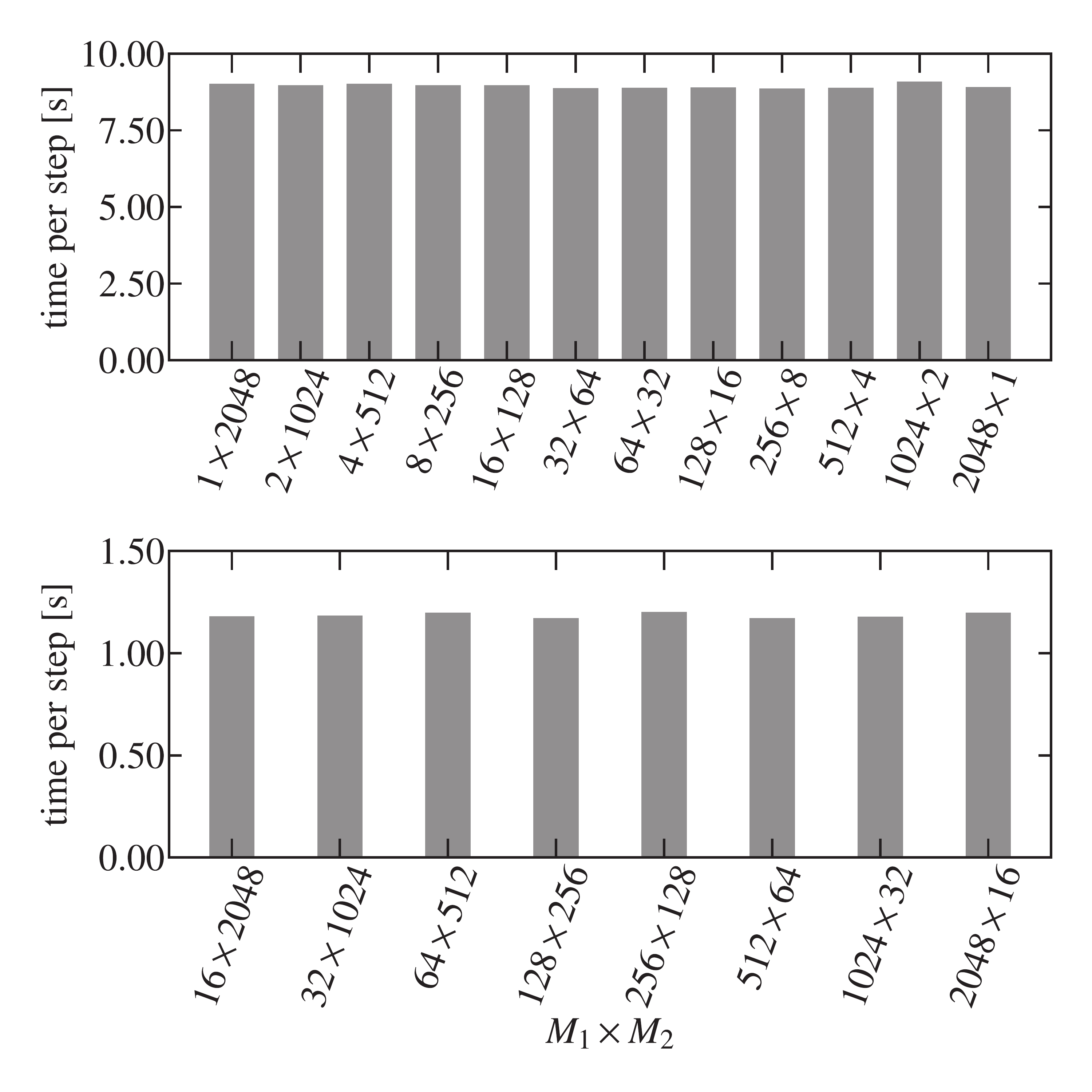}
  \end{center}
  \caption{Performance dependence on the aspect ratio of the MPI process grid at Oakforest-PACS. The setting is the same as in Fig.~\ref{f:scaling}. The total number of MPI processes is fixed to (top) $N_\mr{proc} = 2048$ and (bottom) $N_\mr{proc} = 32768$.}
  \label{f:aspect_ratio}
\end{figure}

In Fig.~\ref{f:scaling}, we chose the aspect ratio of the MPI process grid so that $M_1$ and $M_2$ become as nearly equal as possible.
However, this choice is not always optimal [for example, \citet[Fig.~3]{Pekurovsky2012} showed up to 1.44 times performance difference depending on the aspect ratio].
Thus, we measured the performance of \calliope~at Oakforest-PACS changing the aspect ratio.
The setting was the same as in Fig.~\ref{f:scaling} (i.e., grid number is fixed to $2048^3$, and the thread number is fixed to $N_\mr{thread} = 4$).
As shown in Fig.~\ref{f:aspect_ratio}, we can barely find the performance difference between the choices of the aspect ratio.
Note, however, that the effect of the aspect ratio is supposed to depend on the platform. 

\section{Summary} \label{s:summary}
This paper has introduced a newly developed open-source code, \calliope, which simulates MHD turbulence with high-order accuracy using the pseudospectral method.
\calliope~adopts the P3DFFT library to improve the parallel performance of the three-dimensional FFT computation. 
Due to the pencil decomposition of P3DFFT, almost ideal parallel performance was demonstrated up to $2\times10^5$ processes on the Fugaku supercomputer for $2048^3$ grids. 
We also presented various linear and nonlinear tests to validate the code.
In particular, \calliope~demonstrated a very narrow dissipation range in nonlinear tests, which is the principal merit of the pseudospectral method. 
We highlight that this paper has presented the first simulation of MRI-driven turbulence in isothermal compressible MHD using the pseudospectral method. 
It is anticipated that further interesting properties of MRI-driven turbulence can be identified by analyzing the data obtained in this test. 

Moreover, the inertial range in MRI-driven turbulence may be reached using \calliope~with increased computational resources. 
Revealing the nature of the inertial range is crucial for understanding hot accretion disks. 
For example, it is vital in understanding energy partitioning between ions and electrons~\citep{Kawazura2019,Kawazura2020} and the acceleration of non-thermal particles~\citep{Kimura2016,Sun2021}. 
Finally, although this code was developed to simulate MRI-driven turbulence in shearing coordinates, we anticipate that \calliope~will be a useful tool for studying other turbulence problems.
For example, one can easily modify the RRMHD modules of \calliope~to solve Hall RMHD~\citep[eqs. (5.14)-(5.17) in ][]{Schekochihin2019}, which is a useful model to study turbulence in a regime of electron beta $\sim$ 1 and ion beta $\ll 1$.
We hope that this code will be widely used in the future.

\begin{acknowledgments}
This work was supported by JSPS KAKENHI grant JP19K23451 and JP20K14509. 
The numerical computations reported here were carried out on Fugaku at RIKEN, on Cray XC50 at the Center for Computational Astrophysics at the National Astronomical Observatory of Japan, on ITO at Kyushu University, on Oakforest-PACS and Oakbridge-CX at the University of Tokyo, and on AOBA-B at Tohoku University.
\end{acknowledgments}


\bibliographystyle{yahapj}
\bibliography{references}

\begin{thebibliography}{}
\providecommand\natexlab[1]{#1}
\providecommand\JournalTitle[1]{#1}

\bibitem[{{Balbus} \& {Hawley}(1991)}]{Balbus1991}
{Balbus}, S.~A., \& {Hawley}, J.~F. 1991,
  \href{http://dx.doi.org/10.1086/170270}{\JournalTitle{\apj}, 376, 214}

\bibitem[{{Balbus} \& {Hawley}(1998)}]{Balbus1998}
---. 1998, \href{http://dx.doi.org/10.1103/RevModPhys.70.1}{\JournalTitle{Rev.\
  Mod.\ Phys.}, 70, 1}

\bibitem[{{Biskamp} \& {Schwarz}(2001)}]{Biskamp2001}
{Biskamp}, D., \& {Schwarz}, E. 2001,
  \href{http://dx.doi.org/10.1063/1.1377611}{\JournalTitle{Physics of Plasmas},
  8, 3282}

\bibitem[{{Bodo} {et~al.}(2008){Bodo}, {Mignone}, {Cattaneo}, {Rossi}, \&
  {Ferrari}}]{Bodo2008}
{Bodo}, G., {Mignone}, A., {Cattaneo}, F., {Rossi}, P., \& {Ferrari}, A. 2008,
  \href{http://dx.doi.org/10.1051/0004-6361:200809730}{\JournalTitle{\aap},
  487, 1}

\bibitem[{{Boldyrev}(2006)}]{Boldyrev2006}
{Boldyrev}, S. 2006,
  \href{http://dx.doi.org/10.1103/PhysRevLett.96.115002}{\JournalTitle{\prl},
  96, 115002}

\bibitem[{{Brucker} {et~al.}(2007){Brucker}, {Isaza}, {Vaithianathan}, \&
  {Collins}}]{Brucker2007}
{Brucker}, K.~A., {Isaza}, J.~C., {Vaithianathan}, T., \& {Collins}, L.~R.
  2007,
  \href{http://dx.doi.org/10.1016/j.jcp.2006.10.018}{\JournalTitle{Journal of
  Computational Physics}, 225, 20}

\bibitem[{{Chael} {et~al.}(2018){Chael}, {Rowan}, {Narayan}, {Johnson}, \&
  {Sironi}}]{Chael2018}
{Chael}, A., {Rowan}, M., {Narayan}, R., {Johnson}, M., \& {Sironi}, L. 2018,
  \href{http://dx.doi.org/10.1093/mnras/sty1261}{\JournalTitle{\mnras}, 478,
  5209}

\bibitem[{{Chen} {et~al.}(2011){Chen}, {Mallet}, {Yousef}, {Schekochihin}, \&
  {Horbury}}]{Chen2011}
{Chen}, C.~H.~K., {Mallet}, A., {Yousef}, T.~A., {Schekochihin}, A.~A., \&
  {Horbury}, T.~S. 2011,
  \href{http://dx.doi.org/10.1111/j.1365-2966.2011.18933.x}{\JournalTitle{\mnras},
  415, 3219}

\bibitem[{Czechowski {et~al.}(2012)Czechowski, Battaglino, McClanahan, Iyer,
  Yeung, \& Vuduc}]{Czechowski2012}
Czechowski, K., Battaglino, C., McClanahan, C., {et~al.} 2012, in Proceedings
  of the 26th ACM international conference on Supercomputing, 205

\bibitem[{{Goedbloed} \& {Poedts}(2004)}]{Goedbloed2004}
{Goedbloed}, J.~P.~H., \& {Poedts}, S. 2004, {Principles of
  Magnetohydrodynamics}

\bibitem[{{Goldreich} \& {Sridhar}(1995)}]{Goldreich1995}
{Goldreich}, P., \& {Sridhar}, S. 1995,
  \href{http://dx.doi.org/10.1086/175121}{\JournalTitle{\apj}, 438, 763}

\bibitem[{{Hawley}(2000)}]{Hawley2000}
{Hawley}, J.~F. 2000,
  \href{http://dx.doi.org/10.1086/308180}{\JournalTitle{\apj}, 528, 462}

\bibitem[{{Hawley} {et~al.}(1995){Hawley}, {Gammie}, \& {Balbus}}]{Hawley1995}
{Hawley}, J.~F., {Gammie}, C.~F., \& {Balbus}, S.~A. 1995,
  \href{http://dx.doi.org/10.1086/175311}{\JournalTitle{\apj}, 440, 742}

\bibitem[{{Heinemann} \& {Papaloizou}(2009)}]{Heinemann2009}
{Heinemann}, T., \& {Papaloizou}, J.~C.~B. 2009,
  \href{http://dx.doi.org/10.1111/j.1365-2966.2009.14800.x}{\JournalTitle{\mnras},
  397, 64}

\bibitem[{{Hirai} {et~al.}(2018){Hirai}, {Katoh}, {Terada}, \&
  {Kawai}}]{Hirai2018}
{Hirai}, K., {Katoh}, Y., {Terada}, N., \& {Kawai}, S. 2018,
  \href{http://dx.doi.org/10.3847/1538-4357/aaa5b2}{\JournalTitle{\apj}, 853,
  174}

\bibitem[{{Hoshino}(2015)}]{Hoshino2015}
{Hoshino}, M. 2015,
  \href{http://dx.doi.org/10.1103/PhysRevLett.114.061101}{\JournalTitle{\prl},
  114, 061101}

\bibitem[{{Hosking} \& {Schekochihin}(2020)}]{Hosking2020}
{Hosking}, D.~N., \& {Schekochihin}, A.~A. 2020, \JournalTitle{arXiv e-prints},
  arXiv:2012.01393

\bibitem[{{Hussaini} \& {Zang}(1987)}]{Hussaini1987}
{Hussaini}, M.~Y., \& {Zang}, T.~A. 1987,
  \href{http://dx.doi.org/10.1146/annurev.fl.19.010187.002011}{\JournalTitle{Annual
  Review of Fluid Mechanics}, 19, 339}

\bibitem[{{Iroshnikov}(1963)}]{Iroshnikov1963}
{Iroshnikov}, P.~S. 1963, \JournalTitle{Astron.\ Zh.}, 40, 742

\bibitem[{{Karniadakis} {et~al.}(1991){Karniadakis}, {Orszag}, \&
  {Israeli}}]{Karniadakis1991}
{Karniadakis}, G.~E., {Orszag}, S.~A., \& {Israeli}, M. 1991,
  \href{http://dx.doi.org/10.1016/0021-9991(91)90007-8}{\JournalTitle{J.\
  Comp.\ Phys.}, 97, 414}

\bibitem[{{Kawazura} \& {Barnes}(2018)}]{Kawazura2018}
{Kawazura}, Y., \& {Barnes}, M. 2018,
  \href{http://dx.doi.org/10.1016/j.jcp.2018.01.026}{\JournalTitle{Journal of
  Computational Physics}, 360, 57}

\bibitem[{{Kawazura} {et~al.}(2019){Kawazura}, {Barnes}, \&
  {Schekochihin}}]{Kawazura2019}
{Kawazura}, Y., {Barnes}, M., \& {Schekochihin}, A.~A. 2019,
  \href{http://dx.doi.org/10.1073/pnas.1812491116}{\JournalTitle{Proc.\ Nat.\
  Acad.\ Sci.}, 116, 771}

\bibitem[{{Kawazura} {et~al.}(2021){Kawazura}, {Schekochihin}, {Barnes},
  {Dorland}, \& {Balbus}}]{Kawazura2021}
{Kawazura}, Y., {Schekochihin}, A.~A., {Barnes}, M., {Dorland}, W., \&
  {Balbus}, S.~A. 2021, \href{http://arxiv.org/abs/2110.12434}{{\sffamily
  arXiv:2110.12434 [astro-ph.HE]}}

\bibitem[{{Kawazura} {et~al.}(2020){Kawazura}, {Schekochihin}, {Barnes},
  {TenBarge}, {Tong}, {Klein}, \& {Dorland}}]{Kawazura2020}
{Kawazura}, Y., {Schekochihin}, A.~A., {Barnes}, M., {et~al.} 2020,
  \href{http://dx.doi.org/10.1103/PhysRevX.10.041050}{\JournalTitle{Phys.\
  Rev.\ X}, 10, 041050}

\bibitem[{{Kempski} {et~al.}(2019){Kempski}, {Quataert}, {Squire}, \&
  {Kunz}}]{Kempski2019}
{Kempski}, P., {Quataert}, E., {Squire}, J., \& {Kunz}, M.~W. 2019,
  \href{http://dx.doi.org/10.1093/mnras/stz1111}{\JournalTitle{\mnras}, 486,
  4013}

\bibitem[{{Kimura} {et~al.}(2016){Kimura}, {Toma}, {Suzuki}, \&
  {Inutsuka}}]{Kimura2016}
{Kimura}, S.~S., {Toma}, K., {Suzuki}, T.~K., \& {Inutsuka}, S.-i. 2016,
  \href{http://dx.doi.org/10.3847/0004-637X/822/2/88}{\JournalTitle{\apj}, 822,
  88}

\bibitem[{{Kraichnan}(1965)}]{Kraichnan1965}
{Kraichnan}, R.~H. 1965,
  \href{http://dx.doi.org/10.1063/1.1761412}{\JournalTitle{Phys.\ Fluids}, 8,
  1385}

\bibitem[{{Kunz} \& {Lesur}(2013)}]{Kunz2013}
{Kunz}, M.~W., \& {Lesur}, G. 2013,
  \href{http://dx.doi.org/10.1093/mnras/stt1171}{\JournalTitle{\mnras}, 434,
  2295}

\bibitem[{{Kunz} {et~al.}(2016){Kunz}, {Stone}, \& {Quataert}}]{Kunz2016}
{Kunz}, M.~W., {Stone}, J.~M., \& {Quataert}, E. 2016,
  \href{http://dx.doi.org/10.1103/PhysRevLett.117.235101}{\JournalTitle{\prl},
  117, 235101}

\bibitem[{{Lesur} \& {Longaretti}(2007)}]{Lesur2007}
{Lesur}, G., \& {Longaretti}, P.~Y. 2007,
  \href{http://dx.doi.org/10.1111/j.1365-2966.2007.11888.x}{\JournalTitle{\mnras},
  378, 1471}

\bibitem[{{Lesur} \& {Longaretti}(2011)}]{Lesur2011}
---. 2011,
  \href{http://dx.doi.org/10.1051/0004-6361/201015740}{\JournalTitle{\aap},
  528, A17}

\bibitem[{{Lithwick}(2007)}]{Lithwick2007}
{Lithwick}, Y. 2007,
  \href{http://dx.doi.org/10.1086/522074}{\JournalTitle{\apj}, 670, 789}

\bibitem[{{Loureiro} \& {Boldyrev}(2017)}]{Loureiro2017}
{Loureiro}, N.~F., \& {Boldyrev}, S. 2017,
  \href{http://dx.doi.org/10.1103/PhysRevLett.118.245101}{\JournalTitle{\prl},
  118, 245101}

\bibitem[{{Loureiro} {et~al.}(2016){Loureiro}, {Dorland}, {Fazendeiro},
  {Kanekar}, {Mallet}, {Vilelas}, \& {Zocco}}]{Loureiro2016}
{Loureiro}, N.~F., {Dorland}, W., {Fazendeiro}, L., {et~al.} 2016,
  \href{http://dx.doi.org/10.1016/j.cpc.2016.05.004}{\JournalTitle{Computer
  Physics Communications}, 206, 45}

\bibitem[{{Machida} {et~al.}(2000){Machida}, {Hayashi}, \&
  {Matsumoto}}]{Machida2000}
{Machida}, M., {Hayashi}, M.~R., \& {Matsumoto}, R. 2000,
  \href{http://dx.doi.org/10.1086/312553}{\JournalTitle{\apjl}, 532, L67}

\bibitem[{{Mallet} {et~al.}(2017){Mallet}, {Schekochihin}, \&
  {Chandran}}]{Mallet2017}
{Mallet}, A., {Schekochihin}, A.~A., \& {Chandran}, B.~D.~G. 2017,
  \href{http://dx.doi.org/10.1093/mnras/stx670}{\JournalTitle{\mnras}, 468,
  4862}

\bibitem[{Mininni {et~al.}(2011)Mininni, Rosenberg, Reddy, \&
  Pouquet}]{Mininni2011}
Mininni, P.~D., Rosenberg, D., Reddy, R., \& Pouquet, A. 2011,
  \href{http://dx.doi.org/10.1016/j.parco.2011.05.004}{\JournalTitle{Parallel
  Computing}, 37, 316}

\bibitem[{{Numata} {et~al.}(2010){Numata}, {Howes}, {Tatsuno}, {Barnes}, \&
  {Dorland}}]{Numata2010}
{Numata}, R., {Howes}, G.~G., {Tatsuno}, T., {Barnes}, M., \& {Dorland}, W.
  2010,
  \href{http://dx.doi.org/10.1016/j.jcp.2010.09.006}{\JournalTitle{Journal of
  Computational Physics}, 229, 9347}

\bibitem[{{Orszag}(1969)}]{Orszag1969}
{Orszag}, S.~A. 1969,
  \href{http://dx.doi.org/10.1063/1.1692445}{\JournalTitle{Physics of Fluids},
  12, II}

\bibitem[{{Orszag} \& {Tang}(1979)}]{Orszag1979}
{Orszag}, S.~A., \& {Tang}, C.~M. 1979,
  \href{http://dx.doi.org/10.1017/S002211207900210X}{\JournalTitle{Journal of
  Fluid Mechanics}, 90, 129}

\bibitem[{{Parashar} {et~al.}(2009){Parashar}, {Shay}, {Cassak}, \&
  {Matthaeus}}]{Parashar2009}
{Parashar}, T.~N., {Shay}, M.~A., {Cassak}, P.~A., \& {Matthaeus}, W.~H. 2009,
  \href{http://dx.doi.org/10.1063/1.3094062}{\JournalTitle{Physics of Plasmas},
  16, 032310}

\bibitem[{Pekurovsky(2012)}]{Pekurovsky2012}
Pekurovsky, D. 2012, \JournalTitle{SIAM Journal on Scientific Computing}, 34,
  C192

\bibitem[{{Perrone} \& {Latter}(2021{\natexlab{a}})}]{Perrone2021a}
{Perrone}, L.~M., \& {Latter}, H. 2021{\natexlab{a}},
  \href{http://arxiv.org/abs/2110.13918}{{\sffamily arXiv:2110.13918
  [astro-ph.CO]}}

\bibitem[{{Perrone} \& {Latter}(2021{\natexlab{b}})}]{Perrone2021b}
---. 2021{\natexlab{b}}, \href{http://arxiv.org/abs/2110.14696}{{\sffamily
  arXiv:2110.14696 [astro-ph.CO]}}

\bibitem[{{Ressler} {et~al.}(2015){Ressler}, {Tchekhovskoy}, {Quataert},
  {Chandra}, \& {Gammie}}]{Ressler2015}
{Ressler}, S.~M., {Tchekhovskoy}, A., {Quataert}, E., {Chandra}, M., \&
  {Gammie}, C.~F. 2015,
  \href{http://dx.doi.org/10.1093/mnras/stv2084}{\JournalTitle{\mnras}, 454,
  1848}

\bibitem[{{Rogallo}(1981)}]{Rogallo1981}
{Rogallo}, R.~S. 1981, {Numerical experiments in homogeneous turbulence}, NASA
  STI/Recon Technical Report N

\bibitem[{{Sano} {et~al.}(2004){Sano}, {Inutsuka}, {Turner}, \&
  {Stone}}]{Sano2004}
{Sano}, T., {Inutsuka}, S.-i., {Turner}, N.~J., \& {Stone}, J.~M. 2004,
  \href{http://dx.doi.org/10.1086/382184}{\JournalTitle{\apj}, 605, 321}

\bibitem[{{Schekochihin}(2020)}]{Schekochihin2020}
{Schekochihin}, A.~A. 2020, \JournalTitle{arXiv:2010.00699},
  \href{http://arxiv.org/abs/2010.00699}{{\sffamily arXiv:2010.00699
  [physics.plasm-ph]}}

\bibitem[{{Schekochihin} {et~al.}(2009){Schekochihin}, {Cowley}, {Dorland},
  {Hammett}, {Howes}, {Quataert}, \& {Tatsuno}}]{Schekochihin2009}
{Schekochihin}, A.~A., {Cowley}, S.~C., {Dorland}, W., {et~al.} 2009,
  \href{http://dx.doi.org/10.1088/0067-0049/182/1/310}{\JournalTitle{\apjs},
  182, 310}

\bibitem[{{Schekochihin} {et~al.}(2019){Schekochihin}, {Kawazura}, \&
  {Barnes}}]{Schekochihin2019}
{Schekochihin}, A.~A., {Kawazura}, Y., \& {Barnes}, M.~A. 2019,
  \href{http://dx.doi.org/10.1017/S0022377819000345}{\JournalTitle{J.\ Plasma
  Phys.}, 85, 905850303}

\bibitem[{{Sharma} {et~al.}(2006){Sharma}, {Hammett}, {Quataert}, \&
  {Stone}}]{Sharma2006}
{Sharma}, P., {Hammett}, G.~W., {Quataert}, E., \& {Stone}, J.~M. 2006,
  \href{http://dx.doi.org/10.1086/498405}{\JournalTitle{\apj}, 637, 952}

\bibitem[{{S{\k{a}}dowski} {et~al.}(2017){S{\k{a}}dowski}, {Wielgus},
  {Narayan}, {Abarca}, {McKinney}, \& {Chael}}]{Sadowski2017}
{S{\k{a}}dowski}, A., {Wielgus}, M., {Narayan}, R., {et~al.} 2017,
  \href{http://dx.doi.org/10.1093/mnras/stw3116}{\JournalTitle{\mnras}, 466,
  705}

\bibitem[{{Squire} \& {Bhattacharjee}(2015)}]{Squire2015}
{Squire}, J., \& {Bhattacharjee}, A. 2015,
  \href{http://dx.doi.org/10.1103/PhysRevLett.115.175003}{\JournalTitle{\prl},
  115, 175003}

\bibitem[{{Squire} {et~al.}(2020){Squire}, {Chandran}, \&
  {Meyrand}}]{Squire2020}
{Squire}, J., {Chandran}, B.~D.~G., \& {Meyrand}, R. 2020,
  \href{http://dx.doi.org/10.3847/2041-8213/ab74e1}{\JournalTitle{\apjl}, 891,
  L2}

\bibitem[{{St-Onge} {et~al.}(2020){St-Onge}, {Kunz}, {Squire}, \&
  {Schekochihin}}]{St-Onge2020}
{St-Onge}, D.~A., {Kunz}, M.~W., {Squire}, J., \& {Schekochihin}, A.~A. 2020,
  \href{http://dx.doi.org/10.1017/S0022377820000860}{\JournalTitle{J.\ Plasma
  Phys.}, 86, 905860503}

\bibitem[{{Sun} \& {Bai}(2021)}]{Sun2021}
{Sun}, X., \& {Bai}, X.-N. 2021,
  \href{http://dx.doi.org/10.1093/mnras/stab1643}{\JournalTitle{\mnras}, 506,
  1128}

\bibitem[{{Suzuki} \& {Inutsuka}(2014)}]{Suzuki2014}
{Suzuki}, T.~K., \& {Inutsuka}, S.-i. 2014,
  \href{http://dx.doi.org/10.1088/0004-637X/784/2/121}{\JournalTitle{\apj},
  784, 121}

\bibitem[{{Tchekhovskoy} {et~al.}(2011){Tchekhovskoy}, {Narayan}, \&
  {McKinney}}]{Tchekhovskoy2011}
{Tchekhovskoy}, A., {Narayan}, R., \& {McKinney}, J.~C. 2011,
  \href{http://dx.doi.org/10.1111/j.1745-3933.2011.01147.x}{\JournalTitle{\mnras},
  418, L79}

\bibitem[{{Umurhan} \& {Regev}(2004)}]{Umurhan2004}
{Umurhan}, O.~M., \& {Regev}, O. 2004,
  \href{http://dx.doi.org/10.1051/0004-6361:20040573}{\JournalTitle{\aap}, 427,
  855}

\bibitem[{{Walker} \& {Boldyrev}(2017)}]{Walker2017}
{Walker}, J., \& {Boldyrev}, S. 2017,
  \href{http://dx.doi.org/10.1093/mnras/stx1032}{\JournalTitle{\mnras}, 470,
  2653}

\bibitem[{{Walker} {et~al.}(2016){Walker}, {Lesur}, \& {Boldyrev}}]{Walker2016}
{Walker}, J., {Lesur}, G., \& {Boldyrev}, S. 2016,
  \href{http://dx.doi.org/10.1093/mnrasl/slv200}{\JournalTitle{\mnras}, 457,
  L39}

\bibitem[{{Zhdankin} {et~al.}(2017){Zhdankin}, {Walker}, {Boldyrev}, \&
  {Lesur}}]{Zhdankin2017}
{Zhdankin}, V., {Walker}, J., {Boldyrev}, S., \& {Lesur}, G. 2017,
  \href{http://dx.doi.org/10.1093/mnras/stx372}{\JournalTitle{\mnras}, 467,
  3620}

\end{thebibliography}



\end{document}